\documentclass[onecolumn,preprint,preprintnumbers,amsmath,amssymb]{revtex4}

\usepackage{graphicx}% Include figure files
\usepackage{dcolumn}% Align table columns on decimal point
\usepackage{bm}% bold math

\begin{document}
\title{Interaction between soft magnetic and superconducting films}% Force line breaks with \\

\author{Lars Egil Helseth}
\affiliation{Max Planck Institute of Colloids and Interfaces, D-14424, Potsdam, Germany }%

\begin{abstract}
We study theoretically the interaction between soft magnetic and   
superconducting films. It is shown that the vortex induces a magnetization 
distribution in the magnetic film, thus modifying the magnetic field as well as 
the interactions with e.g. Bloch walls. In particular, the field from the vortex 
may change sign close to the core as a result of the vortex-induced magnetization, 
thus resulting in a crossover from attraction to repulsion between vortices and Bloch 
walls. Moreover, we show that by tuning the anisotropy field of 
the magnetic film, one may enhance the 
interaction between Bloch walls and vortices. Finally, we discuss how the structure 
of a Bloch wall in presence of the thin film superconductor can be revealed by 
magneto-optic imaging using an additional magneto-optic indicator. 

\end{abstract}

\pacs{Valid PACS appear here}
\maketitle

\section{Introduction}
For centuries humans have tried to visualize magnetic fields using 
a variety of different devices. Iron particles, magnetic bacterias, Hall 
probes, ferrofluids, inductive coils, Lorentz microscopy and electron holography are just 
a few of these approaches. The last decades this research has been highly motivated by the need to visualize
magnetic fields from superconductors. To this end, magneto-optic methods have 
proven particularly useful. In 1957 P.B. Alers visualized for the first 
time the magnetic field distribution from a superconductor using cerous nitrate 
glycerol\cite{Alers}. A successful development followed in 1968, when Kirchner 
noted that chalcogenide films can be used to visualize these fields with improved 
sensitivity\cite{Kirchner}. The next two decades magneto-optic imaging studies were 
published periodically, mostly concentrating on large scale flux 
penetration\cite{Habermeier}. A major discovery was done in the beginning 
of the 1990's, when it was found that bismuth-substituted ferrite garnet films with 
in-plane magnetization are excellent indicators with very little domain activity (see e.g. 
Ref. \cite{Dorosinskii}).
This discovery triggered a large number of quantitative investigations of the flux 
behavior in high temperature superconductors, magnetic defects and phase 
transitions, microelectronic circuits and magnetic storage 
media\cite{Jooss1,Koblischka,Koblischka1,Johansen,Polyanskii,Helseth1,Helseth2,Shamonin,Jooss}.
Recently the long standing goal of imaging single vortices was achieved\cite{Goa1,Goa2}. 
This achievment opens a new dimension in the study of interactions between superconductors and soft 
magnetic films, and the first experimental results have already been obtained\cite{Goa3}. 
However, to date no theory exists which can explain future experimental results in this field. 
A key question here is how the vortices interact with the soft magnetic films, in particular 
with micromagnetic elements such as Bloch walls. 
To this end, it should be pointed out that several theoretical studies have been concerned with the interaction between 
vortices and magnetic nanostructures (see e.g. Ref. \cite{Helseth4,Erdin,Milosevic} and references therein). However, the fact 
that the vortex may induce a magnetization distribution in the magnetic film has so far been neglected, and these theories are therefore not 
applicable to soft magnetic films. The aim of the current paper is to take the first step towards a theory which can 
explain the interactions between superconducting and soft magnetic films. 

The thickness of available bismuth-substituted ferrite garnet films is today typically $>500$ $nm$. Here we will only analyze the case of magnetic films much thinner than 
the superconducting penetration depth. This is clearly not justified for current garnet films, but should be of interest 
in order to gain some basic understanding of the system. Moreover, in a previous paper it was proposed theoretically that 
reducing the thickness of the magneto-optic films is of importance when optimizing the magneto-optic image\cite{Helseth3}. Therefore, we 
argue that our theory applies to thin films utilizing the Kerr effect, where all the polarization rotation takes place 
within the first few nanometers of the film. Kerr films may very well be a key element in future magneto-optic imaging, in 
particular if suitable films with high magneto-optic sensitivity and low domain activity can be found.

\section{The thin film system}
Consider a thin superconducting film of infinite extent, located at 
z=0 with thickness $d$ much smaller than the penetration depth of the
superconductor. The surface is covered by an uniaxial soft magnetic film with thickness comparable to or smaller than 
that of the superconducting film. That is, we assume that the magnetic film consists 
of surface charges separated from the superconductor by a very thin oxide layer
(of thickness t) to avoid spin diffusion and proximity effects. In 
general, the current density is a sum of the supercurrents and magnetically 
induced currents, which can be expressed through the generalized London 
equation as\cite{Helseth4} 
\begin{equation}
\mbox{\boldmath $\nabla \times J$} = -\frac{1}{\lambda ^{2}} \mbox{\boldmath $H$} + 
\frac{1}{\lambda ^{2} } V (\mbox{\boldmath $\rho$}) \delta (z) \mbox{\boldmath $\hat{e}$}_{z} + 
\mbox{\boldmath $\nabla \times \nabla \times M$}_{V}\,\,\, ,
\label{LA}
\end{equation}
where $\mbox{\boldmath $J$}$ is the current density, $\lambda$ is the penetration depth, $\mbox{\boldmath $H$}$ is the 
magnetic field and $\mbox{\boldmath $\nabla$} \times \mbox{\boldmath $M$}_{V}$ 
is the magnetically induced current. Note that the magnetically induced currents are included as the last term on the 
right-hand side of Eq. \ref{LA}, and are therefore generated in the same plane as the vortex. This is justified since 
we assume that the thicknesses of both the superconducting and magnetic films are smaller than the penetration depth, and 
the magnetic film is located very close to the superconductor. The vortex is aligned in the z 
direction, and its source function $V(\mbox{\boldmath $\rho$})$ is assumed to
be rotationally symmetric. In the case of a Pearl vortex we may set 
$V(\rho) =(\Phi _{0} /\mu _{0}  )\delta (\rho)$, where $\Phi _{0}$ is the flux quantum and 
$\mu _{0}$ the permeability of vacuum. 

The magnetization $\mbox{\boldmath $M$}_{V}$ is the sum of two different contributions, $\mbox{\boldmath $M$}_{I}$ and $\mbox{\boldmath $M$}_{B}$. 
First, the vortex induces a magnetization $\mbox{\boldmath $M$}_{I}$ in the magnetic film, which again gives rise to opposing 
currents in the superconductor. 
Second, there may be 'hard' micromagnetic elements, e.g. prepatterned magnetization distributions or 
Bloch walls, that do not experience a 
significant change in their magnetization distribution $\mbox{\boldmath $M$}_{B}$ upon interaction 
with the vortex.
The vortex-induced magnetization can be found by assuming that the magnetooptic film is a soft magnet 
consisting of a single domain with in-plane magnetization in absence of external magnetic fields. 
Moreover, we assume that the free energy of this domain can be expressed as a sum of the uniaxial anisotropy 
and the demagnetizing energy. The magnetic field from a vortex tilts the magnetization vector out of the plane, 
and it can be shown that for small tilt angles one has \cite{Helseth3} 
\begin{equation}
\mbox{\boldmath $M$}_{I} =\left( M_{I\rho},M_{Iz} \right) \approx \left(M_{s},\frac{M_{s}H_{vz}}{H_{a} +H_{v\rho}} \right) \,\,\, ,
\label{tilt}
\end{equation}
where $M_{I\rho}=\sqrt{M_{Ix}^{2}+M_{Iy}^{2}}$ and $M_{Iz}$ are the in-plane and perpendicular components of the magnetization, 
$H_{v\rho}=\sqrt{H_{vx}^{2} +H_{vy}^{2}}$ and $H_{vz}$ are the in-plane and perpendicular components of the vortex field, and 
$M_{s}$ is the saturation magnetization of the magnetic film. $H_{a}=M_{s} -2K_{u}/\mu_{0}M_{s}$ is the socalled anisotropy field, where 
$K_{u}$ is the anisotropy constant of the magnetic film. Here the easy axis is normal to the film surface, and we neglect 
the cubic anisotropy of the system, which means that the in-plane magnetization direction must be the same as that induced by the vortex field.  
In the current work we will also assume that the anisotropy field is much larger than the in-plane vortex field, thus 
allowing us to write
\begin{equation}
\mbox{\boldmath $M$}_{I} \approx M_{s}\left(1,\frac{H_{vz}}{H_{a}} \right) \,\,\, .
\label{tilt1}
\end{equation}
Upon using Eq. \ref{tilt1}, we neglect the contribution from the exchange energy, which is 
justifiable for sufficiently large $K_{u}$ or small spatial field gradients. 

In order to solve the generalized London equation, we follow the method of Ref. \cite{Helseth4} and obtain
\begin{equation}
H_{z} - 2\lambda _{e} \frac{\partial H_{z}}{\partial z} = 
V (\rho) +d\lambda _{e}\left( \mbox{\boldmath $\nabla \times \nabla \times M$}_{I} \right) _{z} +d\lambda _{e}\left( \mbox{\boldmath $\nabla \times \nabla \times M$}_{B} \right) _{z}
\,\,\, .
\label{Lond}
\end{equation}
We now apply the superposition
principle to separate the contributions from the vortex and the 'hard' magnetic elements. It is 
important to note that our magnetization distributions have no 
volume charges. 
Using Eq. \ref{tilt1}, the vortex part of Eq. \ref{Lond} can be written as
\begin{equation}
H_{vz} - 2\lambda _{e} \frac{\partial H_{vz}}{\partial z} = 
V(\rho ) -\alpha \left(\frac{\partial ^{2}H_{vz}}{\partial x^{2}} + \frac{\partial ^{2}H_{vz}}{\partial y^{2}} \right) \,\,\, ,  \,\,\,  \alpha = \frac{d\lambda _{e}M_{s} }{H_{a} }      \,\,\, ,
\label{vortexsolution}
\end{equation}
whereas the magnetic part is
\begin{equation}
H_{mz} - 2\lambda _{e} \frac{\partial H_{mz}}{\partial z} = d\lambda _{e}\left( \mbox{\boldmath $\nabla \times
\nabla \times M$}_{B} \right) _{z}
\,\,\, .
\label{magneticsolution}
\end{equation}
Here we define $H_{m\rho}=\sqrt{H_{mx}^{2} +H_{my}^{2}}$ and $H_{mz}$ to be the radial and perpendicular components of
the magnetic field induced by the 'hard' micromagnetic elements.
 
\subsection{Vortex solution}
In order to solve Eq. (\ref{vortexsolution}), we use the method of Ref. \cite{Helseth4}, and 
find the following $z$ and radial components of the magnetic field:
\begin{equation}
H_{vz} (\rho,z) =\frac{1}{2\pi }\int_{0}^{\infty} 
kV(k)\frac{J_{0} (k\rho)}{1+2\lambda _{e}k -\alpha k^{2}} \exp(-k|z|)dk
\,\,\, ,
\end{equation}
\begin{equation}
H_{v\rho} (\rho,z) =\frac{1}{2\pi }\int_{0}^{\infty} 
kV(k)\frac{J_{1} (k\rho)}{1+2\lambda _{e}k -\alpha k^{2}} \exp(-k|z|)dk
\,\,\, .
\end{equation}
In the case $\alpha =0$ and $V(k) =\Phi_{0}/\mu_{0}$, the 
field is reduced to that of the standard Pearl solution\cite{Helseth4,Pearl}.
We also notice that there is a divergency in k-space when $1+2\lambda _{e}k -\alpha k^{2} =0$. This is due to 
the fact that we assumed a very thin superconductor ($d\ll\lambda$). The divergency will therefore smooth out upon 
solving the London equation for arbitrary thicknesses $d$, but this task is outside the scope of the current work.
Nonetheless, it is seen that at 
small distances (large $k$) the magnetic field may change sign as 
compared to the standard Pearl solution, and this is attributed to the magnetically induced currents. 
On the other hand, at large distances (small $k$) the field is basically not influenced by the soft magnetic 
film. A typical scale for the crossover is $k\sim H_{a}/M_{s}d$. One can therefore not expect to reveal this 
phenomenon by magneto-optic imaging, since the length scales are typically much smaller than the wavelength of light.

\subsection{Interactions}
It is of considerable interest to find the interaction between a micromagnetic element with perpendicular 
magnetization $M_{B}\hat{\mbox{\boldmath $e$}}_{z}\delta (z)$ and a vortex. To this end, it should be noted 
that the magnetic induction in the magnetic film is given by 
$\mbox{\boldmath $B$}_{I} = \mu _{0}(\mbox{\boldmath $H$}_{v} +\mbox{\boldmath $M$}_{I})$. 
Here the interaction energy between 
the vortex and the magnetization is given by
\begin{equation}
E_{vm} = -\int_{-\infty}^{\infty} \int_{-\infty}^{\infty} \mbox{\boldmath $M$}_{B} \mbox{\boldmath $\cdot$}\mbox{\boldmath $B_{I}$} d^{2}\rho = -\mu _{0} \int_{-\infty}^{\infty} \int_{-\infty}^{\infty} M_{B}
\hat{\mbox{\boldmath $e$}}_{z} \mbox{\boldmath $\cdot$} \left( \mbox{\boldmath $H_{v}$} + \mbox{\boldmath $M_{I}$} \right) d ^{2}\rho \,\,\, ,
\label{energy}
\end{equation}
where the vortex is assumed to be displaced a distance $\rho _{0}$ from the origin.
It is clear that the radial component of the vortex-induced magnetization does not interact with $\mbox{\boldmath $M$}_{B}$, since the two vectors 
are perpendicular to each other. However, $\mbox{\boldmath $M$}_{B}$ does interact with the $z$-component of the vortex induced field, and we find
\begin{equation}
E_{vm} =  -\mu _{0}\left( 1 + \frac{M_{s}}{H_{a}} \right) \int_{-\infty}^{\infty} \int_{-\infty}^{\infty} 
M_{B} H_{vz} d ^{2}\rho \,\,\, .
\label{interact}
\end{equation}
Equation \ref{interact} can be transformed by using the following formula:
\begin{equation}
\int _{-\infty}^{\infty} \int_{-\infty}^{\infty}A(\rho)B(\rho) d ^{2}\rho 
=\frac{1}{(2\pi )^{2}} \int_{-\infty}^{\infty} \int _{-\infty}^{\infty}
A(-k)B(k) d^{2}k\,\,\, .
\label{help1}
\end{equation}
Then, in the cylindrically symmetric situation one finds
\begin{equation}
E_{vm} =  -\frac{\mu _{0}}{2\pi} \left( 1 + \frac{M_{s}}{H_{a}} \right)  \int_{0}^{\infty} kV(k)\frac{M_{B}(k) 
J_{0} (k\rho _{0}) }{1+2\lambda _{e}k -\alpha k^{2}} dk  \,\,\, .
\label{energyvm}
\end{equation}
The associated force can be found by taking
the derivative with respect to $\rho _{0}$, resulting in
\begin{equation}
F_{vm} = -\frac{\mu _{0}}{2\pi} \left( 1 + \frac{M_{s}}{H_{a}} \right) \int_{0}^{\infty} k^{2}V(k)\frac{M_{B}(k) 
J_{1} (k\rho _{0}) }{1+2\lambda _{e}k -\alpha k^{2}} dk     \,\,\, .
\label{forcevm}
\end{equation}
At small distances (large $k$) the force may change sign due to the magnetically induced currents. Let us now assume that $M_{B}(k)=M_0$, i.e. that 
the magnetization distribution can be represented by a delta function, in order to gain some insight into this phenomenom.
Then, for a Pearl vortex, the force becomes
\begin{equation}
F_{vm} = -\frac{\Phi _{0}M_0}{2\pi} \left( 1 + \frac{M_{s}}{H_{a}} \right) \int_{0}^{\infty} k^{2}\frac{ 
J_{1} (k\rho _{0}) }{1+2\lambda _{e}k -\alpha k^{2}} dk     \,\,\, .
\end{equation}
When the distance is small (still larger than the coherence length $\xi$), we may approximate the force by
\begin{equation}
F_{vm} = \frac{\Phi _{0} M_0}{2\pi \alpha \rho _{0}} \left( 1 + \frac{M_{s}}{H_{a}} \right) \,\,\, , \,\,\, \xi < \rho \ll \frac{M_{s}}{H_{a}} d \,\,\, .
\end{equation}
Surprisingly, we see that the force is repulsive, which can be explained by the magnetically induced currents, which generate 
a magnetic field that opposes that produced by the vortex. 
For comparison, we note that in the case of an infinitely thin superconductor ($\alpha =0$) the force between a magnetic 
element and the vortex is at small distances approximated by
\begin{equation}
F_{vm} = -\frac{\Phi _{0}M_0}{4\pi \lambda _{e}\rho _{0}^{2}} \left( 1 + \frac{M_{s}}{H_{a}} \right) \,\,\, , \,\,\, \xi < \rho \ll \lambda _{e} \,\,\, ,
\end{equation}
which is attractive. 
Attractive behavior also takes place at larger distances $\rho _0 \gg \lambda _e$. 
Then the term containing $\alpha$ becomes unimportant, and the force is therefore always attractive. It is interesting to note that 
the force is enhanced by a factor $1 + M_{s}/H_{a}$ due to the vortex-induced magnetization 
distribution. Therefore it should be possible to tune the interaction by changing the uniaxial anisotropy of 
the magneto-optic film.

In a previous work we showed that the interaction between Bloch walls and vortices in bulk superconductors 
are enhanced due to the image charges generated in the superconducting substrate\cite{Helseth1}. The enhancement 
proposed in the current work has clearly a different origin.

\section{Interaction with Bloch walls}
Magneto-optic films usually contain up to several nearly one dimensional domain walls (in the sense that the 
magnetization within the wall only varies with the $x$-coordinate), and these can interact with the superconductor. 
The $180^{\circ}$ Bloch wall is quite common in 
garnet films, and also in very thin Kerr films studied here, although Neel walls may also occur in thin films if the 
demagnetization energy is significant. Neglecting the demagnetization energy, the magnetization vector distribution 
within a Bloch wall is given by\cite{Craik}
\begin{equation}
\mbox{\boldmath $M$}^{Bloch} =M_{s}\left( cos\theta , sin\theta \right)\,\,\, , \,\,\, \theta =2tan^{-1}\left[ exp\left(\sqrt{\frac{K_{u}}{A}}x \right) \right] \,\,\, ,
\end{equation}
where $A$ is the exchange parameter. In addition to naturally occuring domain walls, it is also possible to 
pattern the magneto-optic film by depositing thin nanomagnets on its surface. 
Here we will model a Bloch wall using the following two different simple models with only a $z$ component:
\begin{equation}
\mbox{\boldmath $M$}^{G} = M_{s} \exp (-\beta x^{2}) \delta(z)\mbox{\boldmath $\hat{e}$}_{z} \,\,\, , 
\label{gauss}
\end{equation}
\begin{displaymath}
\mbox{\boldmath $M$}^{S} = \left\{ \begin{array}{ll}
M_{s}\delta(z)\mbox{\boldmath $\hat{e}$}_{z} & \textrm{if  $-W \leq x \leq
W$}\\
0 & \textrm{if $|x|  \geq W$}\\  
\end{array} \right. ,
\end{displaymath}
where we assume that $W=\sqrt{2/\beta }$. Using the method of Ref. \cite{Helseth4}, we find that the Gaussian magnetization distribution results in the following 
field components:
\begin{equation}
H^{G}_{z}(x,z) =\frac{\lambda _{e} d M_{s}}{\sqrt{\pi \beta} }\int_{0}^{\infty}k_{x} ^{2} \frac{\exp\left( -k_{x}^{2}/4\beta \right) cos(k_{x}x)}{1+2\lambda _{e}k_{x}}
\exp(-k_{x}|z|)dk_{x} \,\,\, ,
\end{equation}
\begin{equation}
H^{G}_{x}(x,z) =\frac{\lambda _{e} d M_{s}}{\sqrt{\pi \beta} }\int_{0}^{\infty}k_{x} ^{2} \frac{ \exp\left( -k_{x}^{2}/4\beta  \right) sin(k_{x}x)}{1+2\lambda _{e}k_{x}}
\exp(-k_{x}|z|)dk_{x}
\,\,\, .
\end{equation}
On the other hand, the step magnetization distribution gives 
\begin{equation}
H^{S}_{z}(x,z) =\frac{2\lambda _{e} d M_{s}}{\pi}\int_{0}^{\infty}k_{x}  \frac{sin(k_{x}W) cos(k_{x}x)}{1+2\lambda _{e}k_{x}}
\exp(-k_{x}|z|)dk_{x} \,\,\, ,
\end{equation}
\begin{equation}
H^{S}_{x}(x,z) =\frac{2\lambda _{e} d M_{s}}{\pi }\int_{0}^{\infty}k_{x}\frac{ sin(k_{x}W) sin(k_{x}x)}{1+2\lambda _{e}k_{x}}
\exp(-k_{x}|z|)dk_{x}
\,\,\, .
\end{equation}

\begin{figure}
\includegraphics[width=12cm]{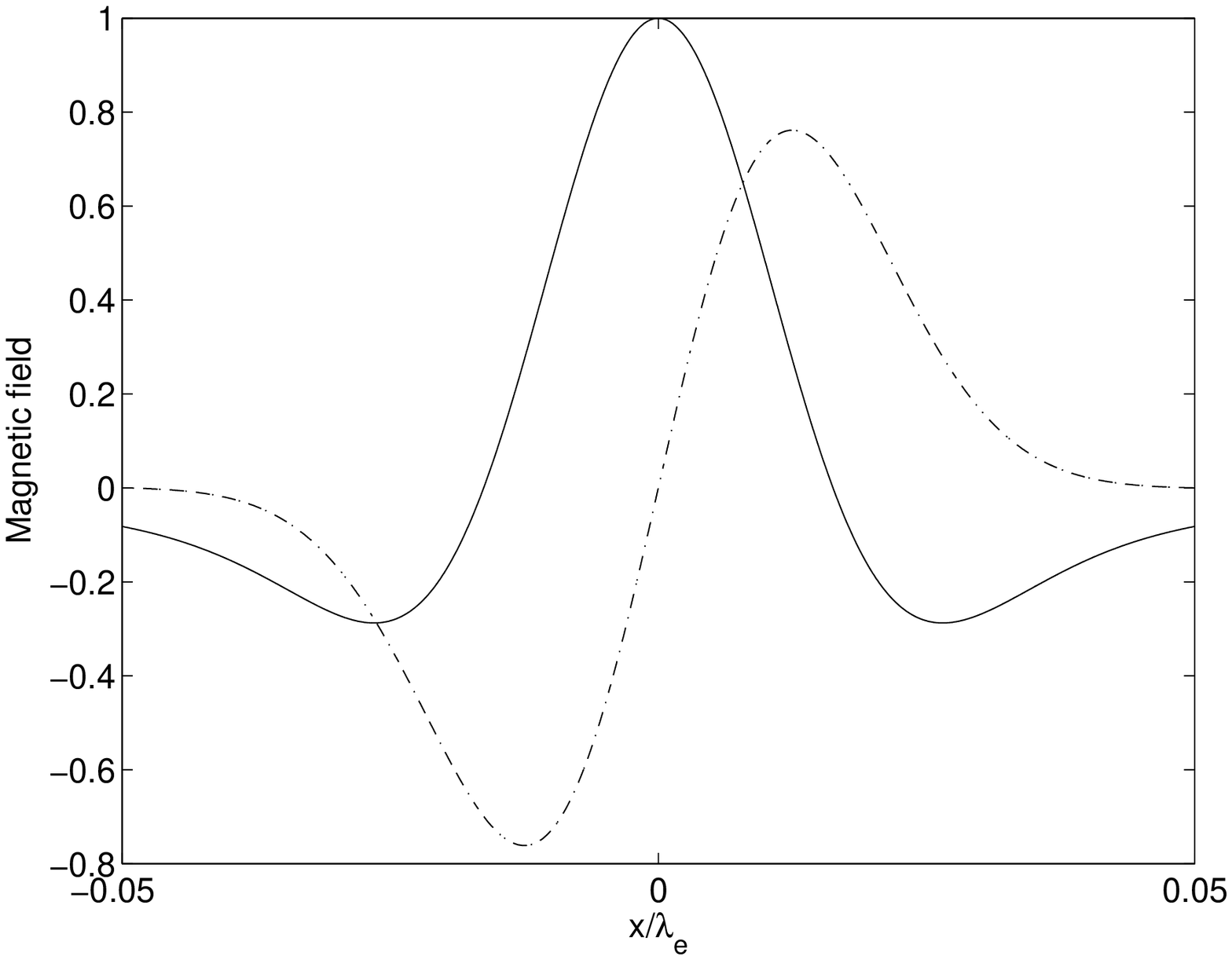}
\caption{\label{f1}  The x (dash-dotted line) and z (solid line) component of the magnetic field generated by a 
gaussian magnetization distribution. Here z=0 and $W=\lambda _{e}/40$. The curves are normalized with
respect to the maximum peak of the z component.}
\vspace{2cm}
\end{figure}

\begin{figure}
\includegraphics[width=12cm]{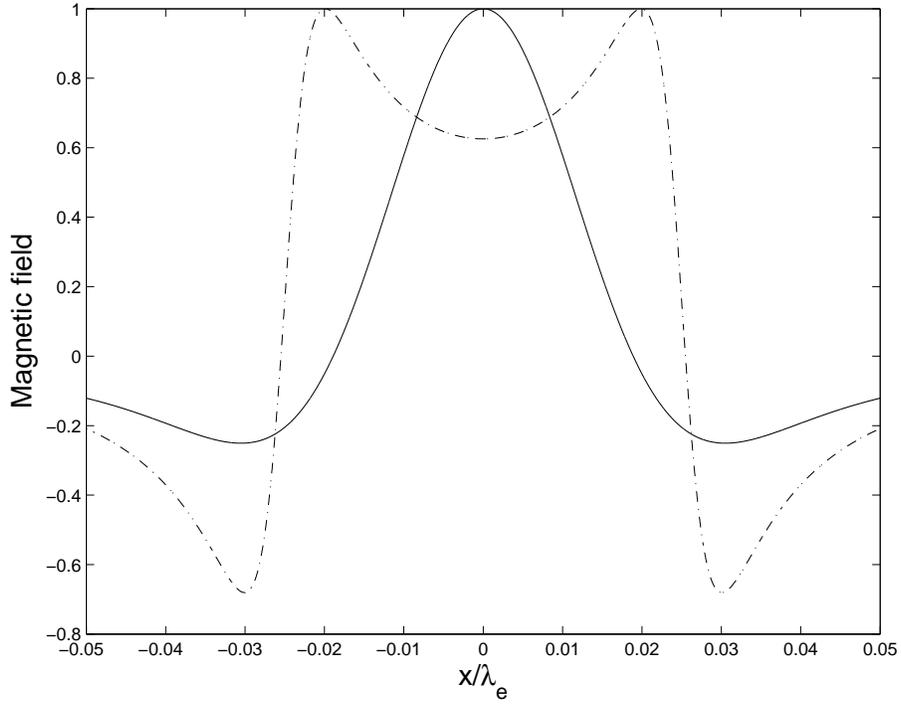}
\caption{\label{f2}  The z components of the magnetic field generated by 
gaussian (solid line) and step-like (dash-dotted line) magnetization distributions. 
Here $z=\lambda_{e}/200$ and $W=\lambda _{e}/40$. The curves are normalized with
respect to the maximum peak of each of the z components.}
\vspace{2cm}
\end{figure}

Figure \ref{f1} shows $H^{G}_{z}$ (solid line) and $H^{G}_{x}$ (dash-dotted line) for $\lambda _{e}=40W$ and $z=0$. 
As expected, the $H^{G}_{z}$ is symmetric about the origin and has its maximum field here, whereas 
$H^{G}_{x}$ is zero at $x=0$. Note also that $H^{G}_{x}$ changes sign at $x=0$, and has peaks at $x/\lambda _{e} \approx \pm 0.02$. Figure \ref{f2} shows 
$H^{G}_{z}$ (solid line) and $H^{S}_{z}$ (dash-dotted line) for $\lambda _{e}=40W$ and $z=\lambda _{e}/200$. 
Note that in contrast to $H^{G}_{z}$, $H^{S}_{z}$ has two peaks at the edges of the magnetization distribution. 
Thus, it is seen that these two magnetization distributions give rise to different characteristic features 
which may be resolved by magneto-optic imaging.

\begin{figure}
\includegraphics[width=12cm]{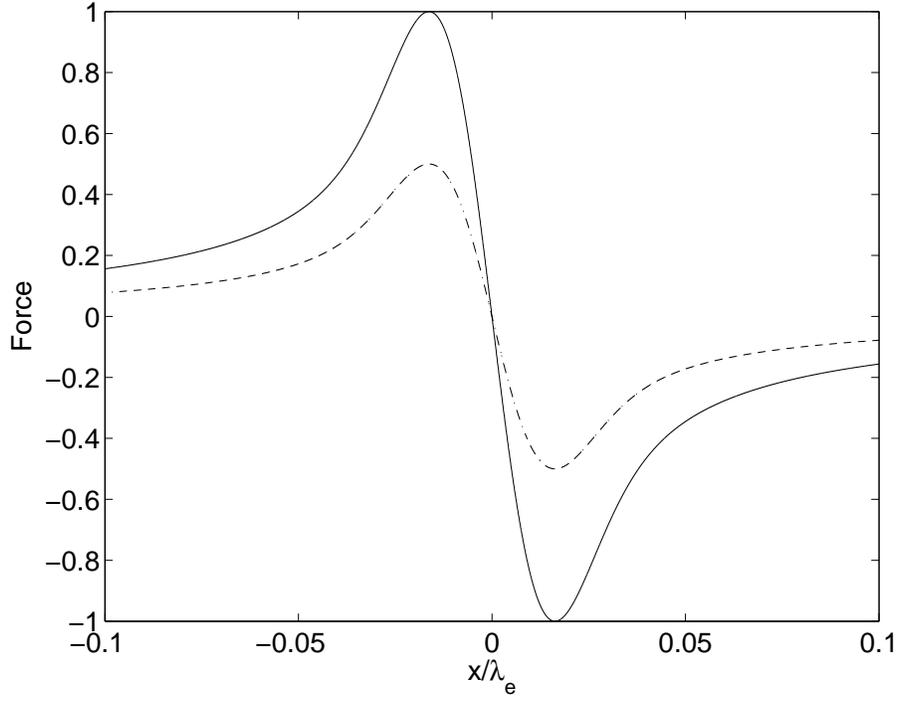}
\caption{\label{f3}  The normalized interaction force for a gaussian magnetization distribution when
$\lambda _{e}=\sqrt{3200/\beta}$. The solid line corresponds to $H_{a}=0.2M_{s}$, and the dash-dotted line 
to $H_{a}=0.5M_{s}$.}
\vspace{2cm}
\end{figure}

The force between the one dimensional magnetization distribution and a Pearl vortex is found to be 
\begin{equation}
F (x) =  -\frac{\Phi _{0}}{\pi} \left( 1 + \frac{M_{s}}{H_{a}} \right) \int_{0}^{\infty} k_{x}\frac{M(k_{x})
sin(k_{x}x)}{1+2\lambda _{e}k_{x} -\alpha k_{x}^{2}} dk_{x}     \,\,\, .
\label{Bloch-vortex}
\end{equation}
If the Bloch wall is represented by a Gaussian magnetization distribution, the interaction force is 
found from Eq. \ref{Bloch-vortex} to be
\begin{equation}
F (x) =  -\frac{M_{s}\Phi _{0}}{\sqrt{\pi \beta }} \left( 1 + \frac{M_{s}}{H_{a}} \right) \int_{0}^{\infty} k_{x}\frac{\exp(-k_{x}^{2}/4\beta )
sin(k_{x}x)}{1+2\lambda _{e}k_{x} -\alpha k_{x}^{2}} dk_{x}     \,\,\, .
\label{Bloch-v}
\end{equation}
Fig. \ref{f3} shows the interaction force in the limit $d\rightarrow 0$. 
The solid line corresponds to $H_{a}=0.2M_{s}$ and the dash-dotted line to $H_{a}=0.5M_{s}$. It is seen 
that the force is always attractive.

\section{Magneto-optic imaging of Bloch walls}
\begin{figure}
\includegraphics[width=12cm]{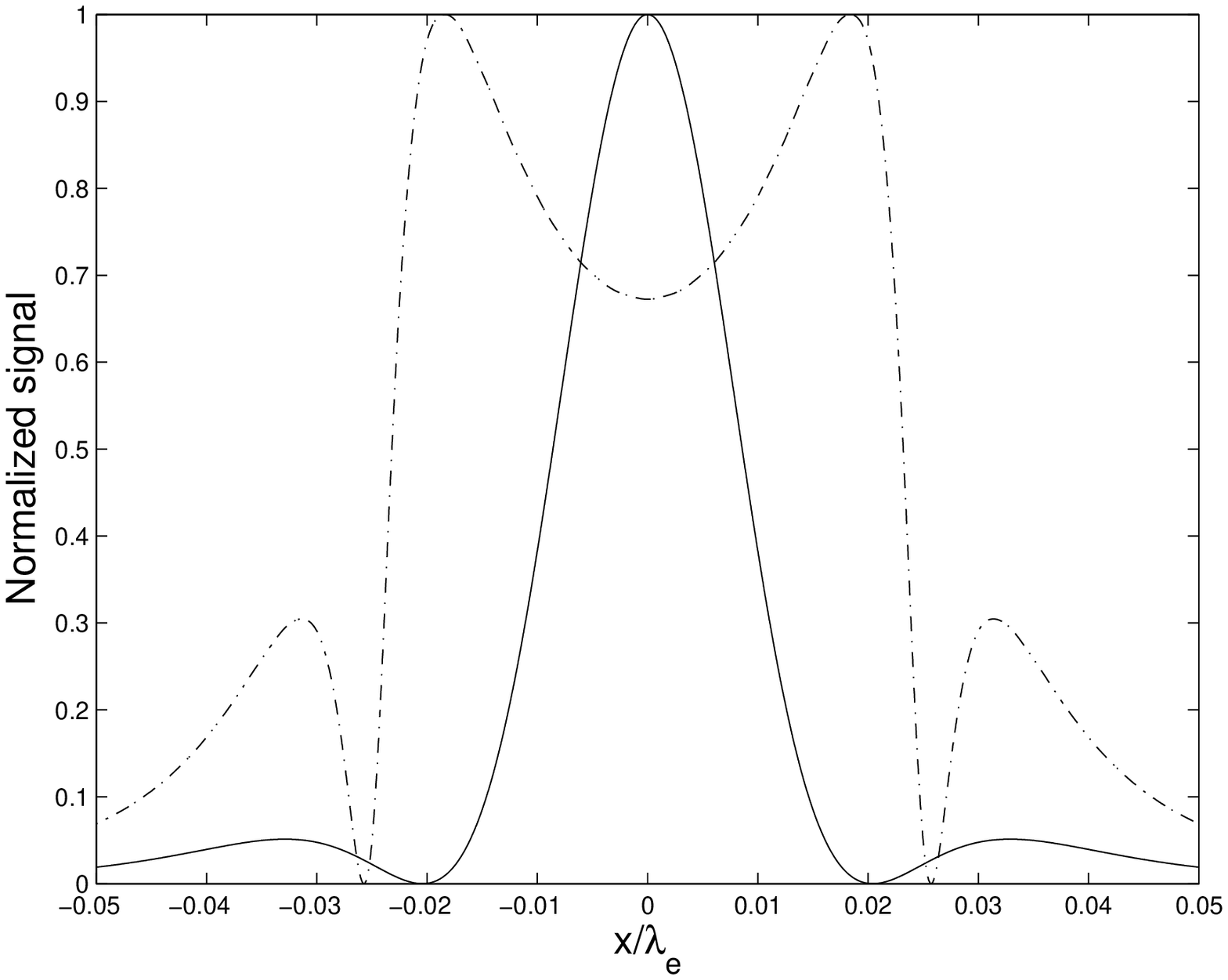}
\caption{\label{f4}  The normalized signal as a function of position for a gaussian (solid line) and 
step-like (dash-dotted line) magnetization distribution. Here we have chosen $W=\lambda _{e} /40$, l=W/10 and
D=W/2.}
\vspace{2cm}
\end{figure}

\begin{figure}
\includegraphics[width=12cm]{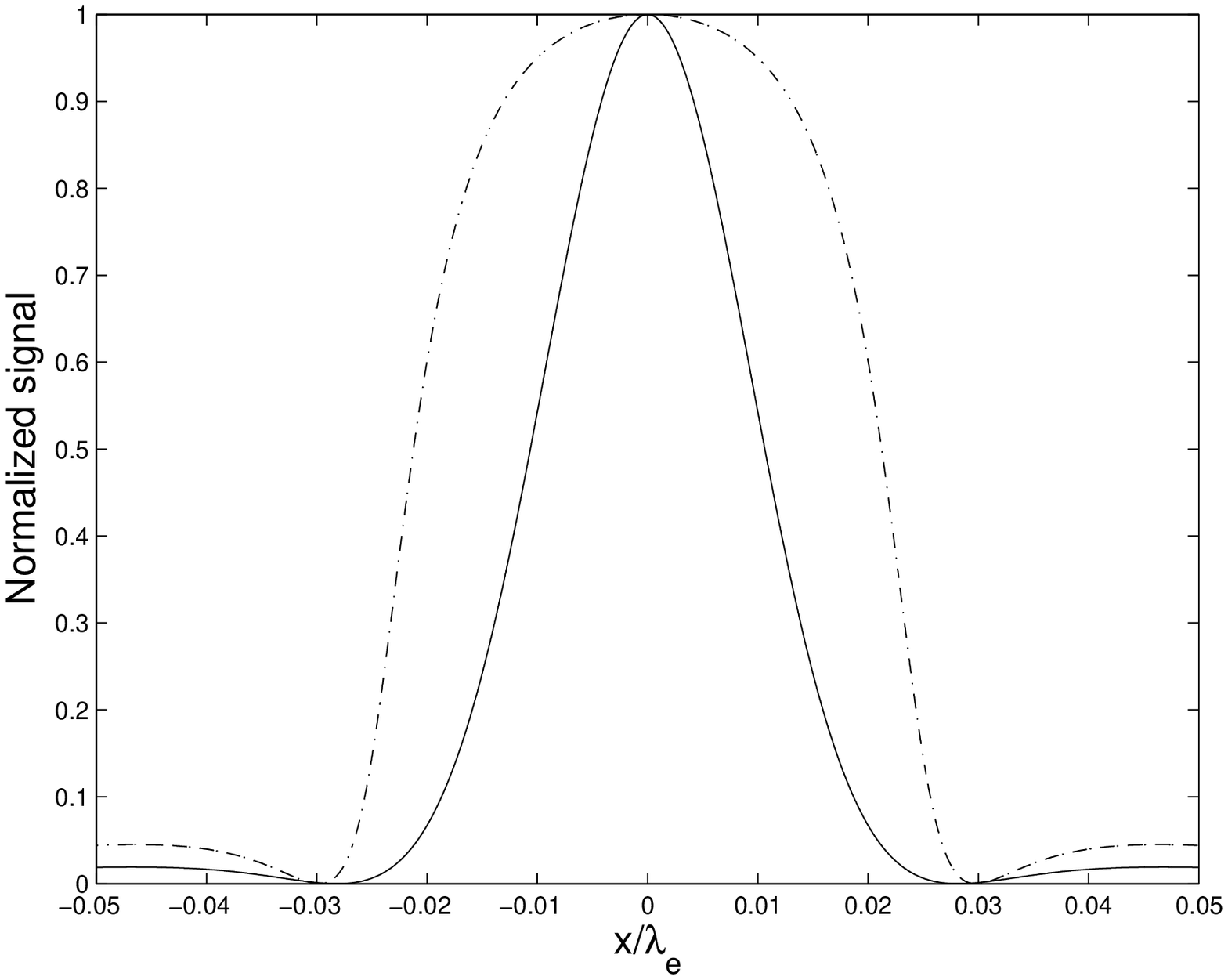}
\caption{\label{f5}  The normalized signal as a function of position for a gaussian (solid line) and 
step-like (dash-dotted line) magnetization distribution. Here we have chosen $W=\lambda _{e} /40$, l=W/5 and
D=2W.}
\vspace{2cm}
\end{figure}

A powerful property of magneto-optic films is that they can visualize the vortices and and Bloch walls simultaneously, thus
allowing us to study the dynamics of these elements. However, it is difficult to draw conclusions about the field from the 
Bloch wall based on a single magneto-optic film, unless the width of the wall is altered by the superconductor\cite{Helseth1}. 
On the other hand, it is of considerable interest to study the structure of the Bloch wall in presence of the
superconducting film in order to gain additional information about the interaction between such systems. This can be 
done by placing a second magneto-optic indicator above our thin films system (composed of a thin superconducting and magneto-optic 
film). However, care must be taken, since the signal detected by the polarization microscope depends on the gap 
$l$ between the thin film system and the second indicator. Moreover, let us assume that the second indicator has thickness $D$, and does not 
interact with the thin film system. A model for magneto-optic imaging was introduced in Ref. \cite{Helseth3}, and this will 
be used here (neglecting absorbtion). 
Figures \ref{f4} and \ref{f5} show the signal modulation for Gaussian and step-like magnetization distributions using two different indicators. In Fig. \ref{f4} a very 
thin indicator (D=W/2, $\lambda _{e}=40W$) positioned close to the thin film 
structure (l=W/10) is used. We see that the characteristic magnetic field profiles can be revealed. However, in Fig. \ref{f5} 
a thicker indicator (D=2W, $\lambda _{e} =40W$) and a larger gap $l$ ($l$=W/5) is assumed. Although the signal profile generated by 
the Gaussian distribution does not change much, it is seen that the characteristic peaks from the step like distribution
are lost. Thus, we conclude that a small gap and a thin indicator is necessary to reveal the characteristic magnetic field profiles 
discussed above. It should be pointed out that vortices may also be present, thus changing the detected signal. However, 
a study of this effect is outside the scope of the current work.

\section{Conclusion}
We have studied the interaction between magnetooptic and   
superconducting films. It was shown that the vortex induces a magnetization 
distribution in the magnetooptic film, thus modifying the magnetic field as well as 
the interaction Bloch walls. In particular, we show that 
by tuning the anisotropy field of the magnetooptic film, one may enhance this 
interaction. We also discussed some simple implications for magnetooptic imaging. 

\acknowledgments
The author is grateful to H. Hauglin, P.E. Goa, M. Baziljevich, T.H. Johansen and E.I. Il'yashenko for many 
interesting discussions on this topic, and also to T.M. Fischer for generous support.

\newpage
\newpage

\end{document}